\documentclass[twocolumn,twoside,slac_two]{revtex4}
\usepackage{graphicx}
\usepackage{fancyhdr}
\usepackage{subfigure}
\usepackage{xspace}
\usepackage{multirow}

\def \babar {\mbox{\sl B\kern-0.1em{\scriptsize A}\kern-0.1em B\kern-0.1em{\scriptsize A\kern-0.2em R}}}
\def \pep2 {PEP-II}
\def \BR {{\ensuremath{\cal B}\xspace}}
\def \epem {\ensuremath{e^+e^-}}
\def \invfb {\ensuremath{\mathrm{fb}^{-1}}}
\def \MeVcc {\ensuremath{\mathrm{MeV}/c^2}}
\def \GeVcc {\ensuremath{\mathrm{GeV}/c^2}}
\def \YOneS {\ensuremath{\Upsilon{(1S)}}}
\def \YTwoS {\ensuremath{\Upsilon{(2S)}}}
\def \YThreeS {\ensuremath{\Upsilon{(3S)}}}
\def \YFourS {\ensuremath{\Upsilon{(4S)}}}
\def \YFiveS {\ensuremath{\Upsilon{(5S)}}}

\begin{document}

\title{Quarkonium States at $B$-Factories}
\author{E. Robutti}
\affiliation{I.N.F.N. - Sezione di Genova, Italy}
\author{from the \babar~ Collaboration}

\begin{abstract}
An overview is given on recent progress in the study of quarkonium spectroscopy at the $B$-factories. In particular, an updated status report is presented of the long list of ``charmonium-like'' resonances newly discovered, whose assignment as true charmonium states is in most cases at least controversial. Also, new measurements on the decay properties of bottomonium states above open-$B$ production thresholds are shown.
\end{abstract}

\maketitle

\thispagestyle{fancy}


\section{Introduction}

Much of the progress attained in recent years in the study of the quarkonium spectra is owed to the measurements performed at $B$-factories. The impressive amount of data recorded by the \babar~ and Belle experiments has made it possible to study rare decay chains and to look for as yet undiscovered resonances in the charmonium and bottomonium mass regions.\\
Results presented here are based on different sub-samples of the full datasets recorded up to now by the two experiments, corresponding to integrated luminosities of about 430 \invfb (\babar~ - final) and about 850 \invfb (Belle). Significant contributions also come from the analysis of the various data samples recorded by the CLEO detector.

\section{The charmonium region}

\subsection{Charmonium production at $B$-factories}

Charmonium states are copiously produced at $B$-factories in a variety of processes.\\
$B$ mesons decay to charmonium in about 3\% of cases.\\
Charmonium can also be produced in the continuum $\epem \rightarrow c \overline{c}$ process: for the $J/\psi$, for instance, the inclusive production cross section is about $2.5~\mathrm{pb}$ in the energy region of the $B$-factories. Of particular interest here is the ``double-charmonium production'' process, first observed by Belle, where a $J/\psi$ is produced together and exclusively with another charmonium state.\\
States with $C = +1$ are formed in $\gamma \gamma$ fusion, where two virtual photons are emitted by the colliding \epem pair ($\epem \rightarrow \epem \gamma^* \gamma^* \rightarrow \epem (c \overline{c})$). About 1 million $\eta_c$ are produced in $100~\invfb$-equivalent of data via this mechanism.\\
Finally, states with $J^{PC} = 1^{--}$ are formed in initial state radiation (ISR) events, where a photon is emitted by the incoming electron or positron ($\epem \rightarrow \gamma (c \overline{c})$). The cross section for ISR production of the $J/\psi$ at the energy of the \YFourS~ peak is about 2 pb.\\

\subsection{The $X(3872)$ state}

The discovery of the narrow  $X(3872)$ state by Belle in 2003 \cite{ref:Belle_X3872_obs} has marked the start of a ``new'' spectroscopy in the charmonium mass region. Since then the resonant structure has been observed in different production and decay channels by CDF , D0, \babar~and again Belle. Besides the discovery production mode $B \rightarrow X(3872) K$, the $X(3872)$ has been observed in prompt $p \overline{p}$ production \cite{ref:CDF-D0_X3872_obs}; searches in prompt \epem~ production \cite{ref:BaBar_X3872_prompt} and in \epem or $\gamma \gamma$ formation \cite{ref:CLEO_X3872_gg-ISR} have given negative results so far. In addition to the discovery decay mode $X(3872) \rightarrow J/\psi~ \pi^+ \pi^-$ (consistent with $X(3872) \rightarrow J/\psi~ \rho^0$), the decay channels $X(3872) \rightarrow J/\psi~ \gamma$ \cite{ref:X3872_Jpsigamma} and $X(3872) \rightarrow D^0 \overline{D}^0 \pi^0$ \cite{ref:X3872_Belle_DDpi} have been seen.\\
Detailed angular analyses \cite{ref:X3872_angular}, together with constraints on quantum numbers from observed production and decay modes, favour the $J^{PC} = 1^{++}$ assignment, although $J^{PC} = 2^{-+}$ is not ruled out.\\
The ``natural'' interpretation of the $X(3872)$ as a charmonium state meets with considerable difficulties \cite{ref:Eichten_review08}: several alternative models have been proposed in recent years. Given the proximity of its mass to the sum of the $D^0$ and $D^{*0}$ masses, the $X(3872)$ might be an $S$-wave $D^0$-$D^{*0}$ bound state, or ``molecule'' \cite{ref:DDstarMolecule}. A different approach interprets the state as a ``tetraquark'' structure, or a bound state of a diquark-antidiquark pair \cite{ref:tetraquarks}. In this case a total of 4 nearly-degenerate states (2 neutral + 2 charged) should exist in the same mass region. Other non conventional interpretations, such as a $c\overline{c}g$ hybrid, are disfavoured, mainly because of the mass, which is inconsistent with current predictions.

\begin{figure*}[!t]
  \centering
    \subfigure[]
      {\includegraphics[width=80mm]{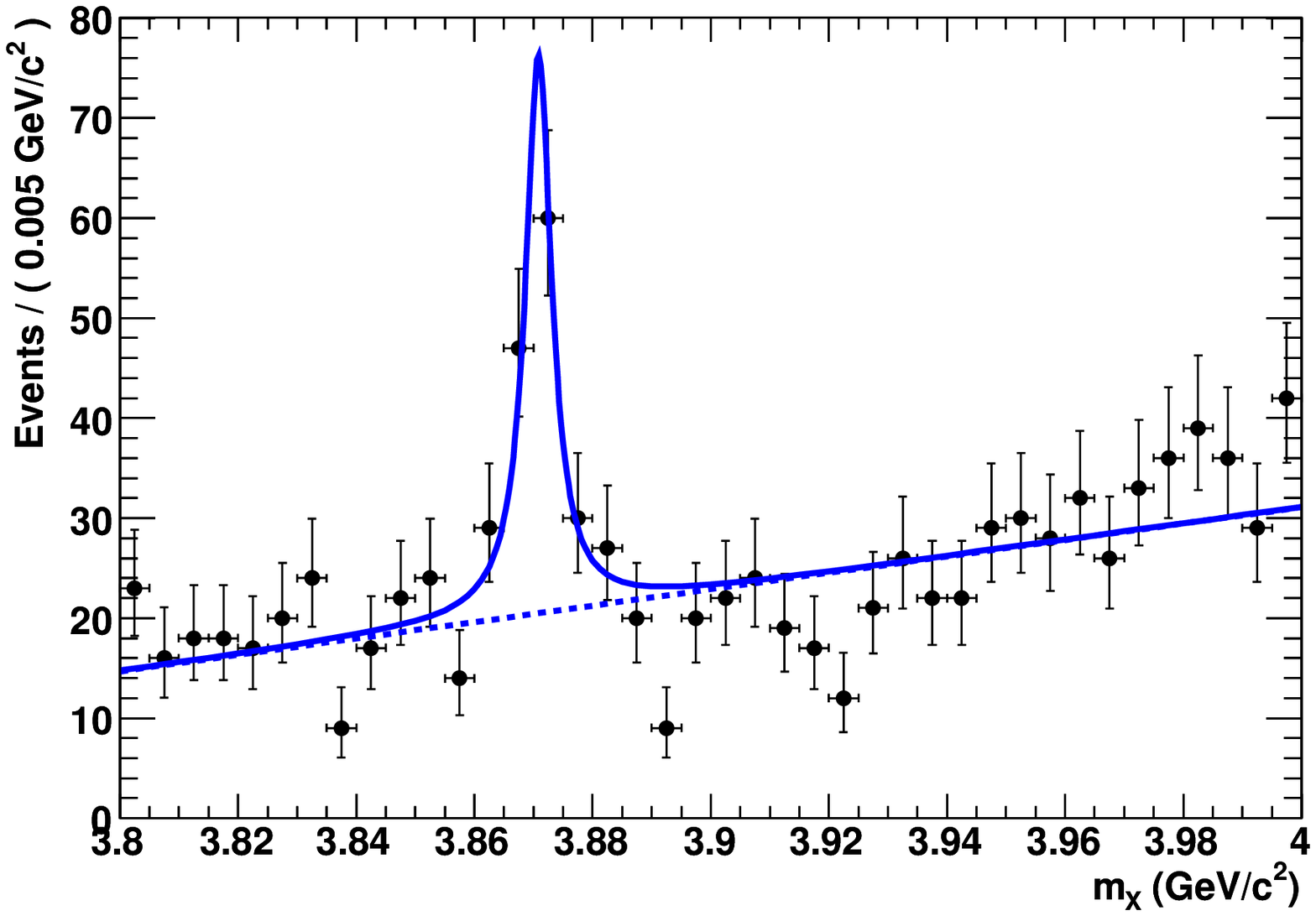}}
    \subfigure[]
      {\includegraphics[width=80mm]{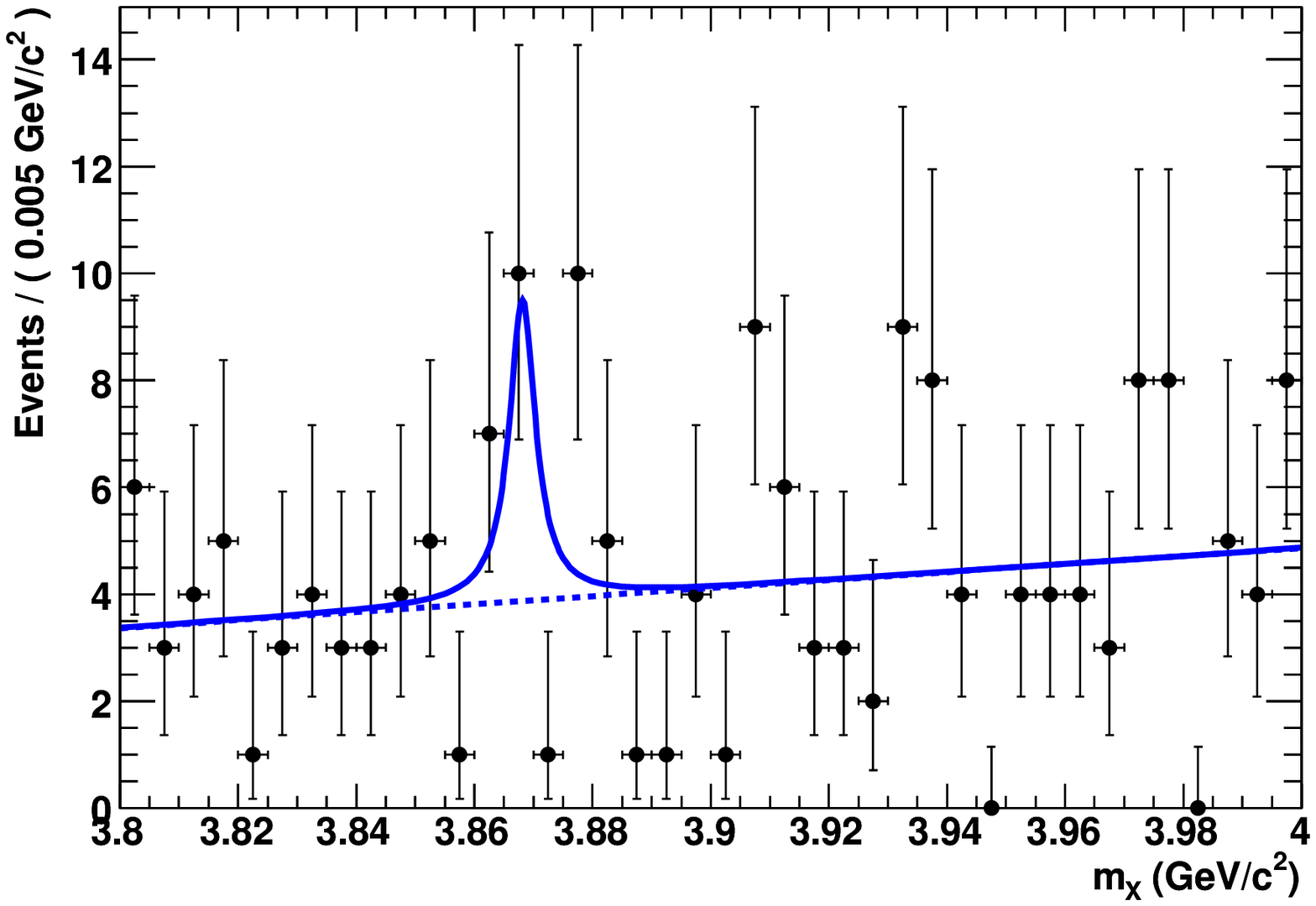}}
    \subfigure[]
      {\includegraphics[width=70mm]{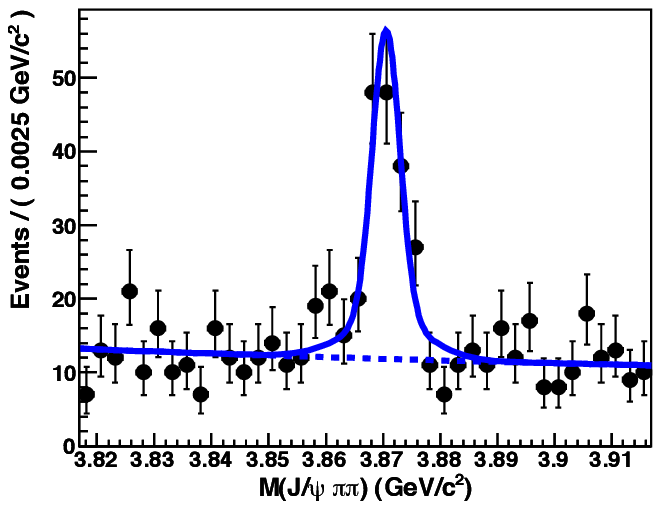}}
    \hspace{8mm}
    \subfigure[]
      {\includegraphics[width=70mm]{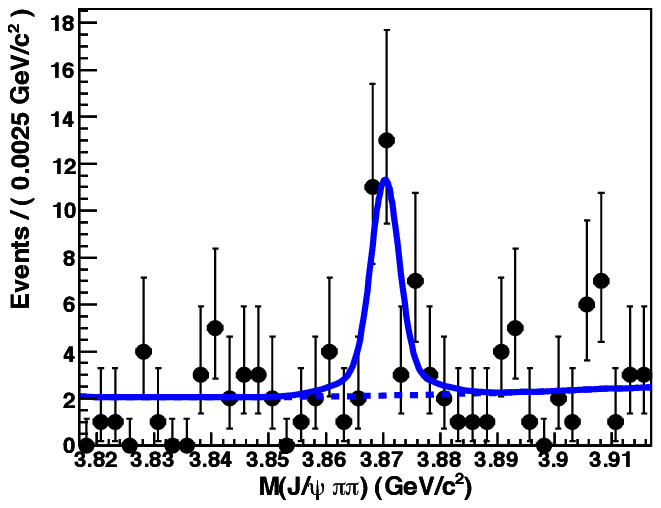}}
    \caption{The $J/\psi~ \pi^+\pi^-$ invariant mass distributions in the $X(3872)$ region for selected $B$ decay events: a) \babar, $B^+ \rightarrow J/\psi~ \pi^+ \pi^- K^+$; b) \babar, $B^0 \rightarrow J/\psi~ \pi^+ \pi^- K^0_\mathrm{S}$; c) Belle, $B^+ \rightarrow J/\psi~ \pi^+ \pi^- K^+$; d) Belle, $B^0 \rightarrow J/\psi~ \pi^+ \pi^- K^0_\mathrm{S}$.}
    \label{fig:X3872-Jpsipipi}
\end{figure*}

In their most recent updates of the $J/\psi~ \pi^+ \pi^-$ analyses, both \babar~ and Belle have shown results for the neutral, as well as for the charged, $B$ decay mode \cite{ref:X3872_Jpsipipi_new}. Using samples corresponding to $413~\invfb$ and $605~\invfb$ respectively, they reconstructed $B^+ \rightarrow J/\psi~ \pi^+ \pi^- K^+$\footnote{Charge-conjugate modes are implied here and in the following.} and $B^0 \rightarrow J/\psi~ \pi^+ \pi^- K^0_\mathrm{S}$ decays (Fig.~\ref{fig:X3872-Jpsipipi}). Belle reports the first statistically significant observation ($5.9 \sigma$) of the production mode $B^0 \rightarrow X(3872) K^0_\mathrm{S}$; weaker evidence is also seen in \babar~ data.\\
The comparison between the charged and neutral channels is of particular interest: in the tetraquark model, two different $X$ states are produced in the two decays, their masses differing by $\sim7\div10~\MeVcc$; moreover, in $D^0$-${D}^{*0}$ molecular models the ratio of the neutral to the charged $B$ decay rate may show departure from unity~\cite{ref:DDstarMolNew}. Table~\ref{tab:X3872-Jpsipipi} reports results for the ratio of branching fractions and for the mass difference $\Delta M \equiv M_{X_\mathrm{H}} - M_{X_\mathrm{L}}$: the ratio appears significantly higher than 0.1, while the measured values of $\Delta M$ are consistent with no mass splitting.

\begin{table}[h]\addtolength{\tabcolsep}{2mm}
  \begin{center}
    \caption{Results from latest $B \rightarrow J/\psi~ \pi^+ \pi^- K$ analyses from \babar~ and Belle.\\}
    \begin{tabular}{l | c @{ $\pm$ } c @{ $\pm$ } c | c @{ $\pm$ } c @{ $\pm$ } c}
      \hline \hline
      & \multicolumn{3}{|c|}{$\frac{\BR(B^0 \rightarrow X(3872) K^0)}{\BR(B^+ \rightarrow X(3872) K^+)}$} & \multicolumn{3}{|c}{$\Delta M (\MeVcc)$} \\
      \hline
      \babar & 0.41 & 0.24 & 0.05 & 2.6 & 1.6 & 0.4 \\
      Belle & 0.82 & 0.22 & 0.05 & 0.18 & 0.89 & 0.26 \\
      \hline
    \end{tabular}
  \label{tab:X3872-Jpsipipi}
  \end{center}
\end{table}

\begin{figure*}[!t]
  \centering
    \subfigure[]
      {\includegraphics[width=72mm]{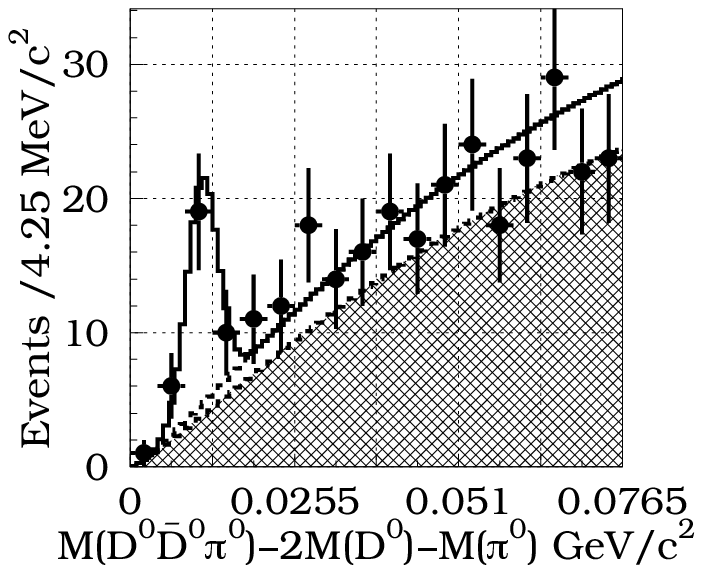}}
    \subfigure[]
      {\includegraphics[width=90mm]{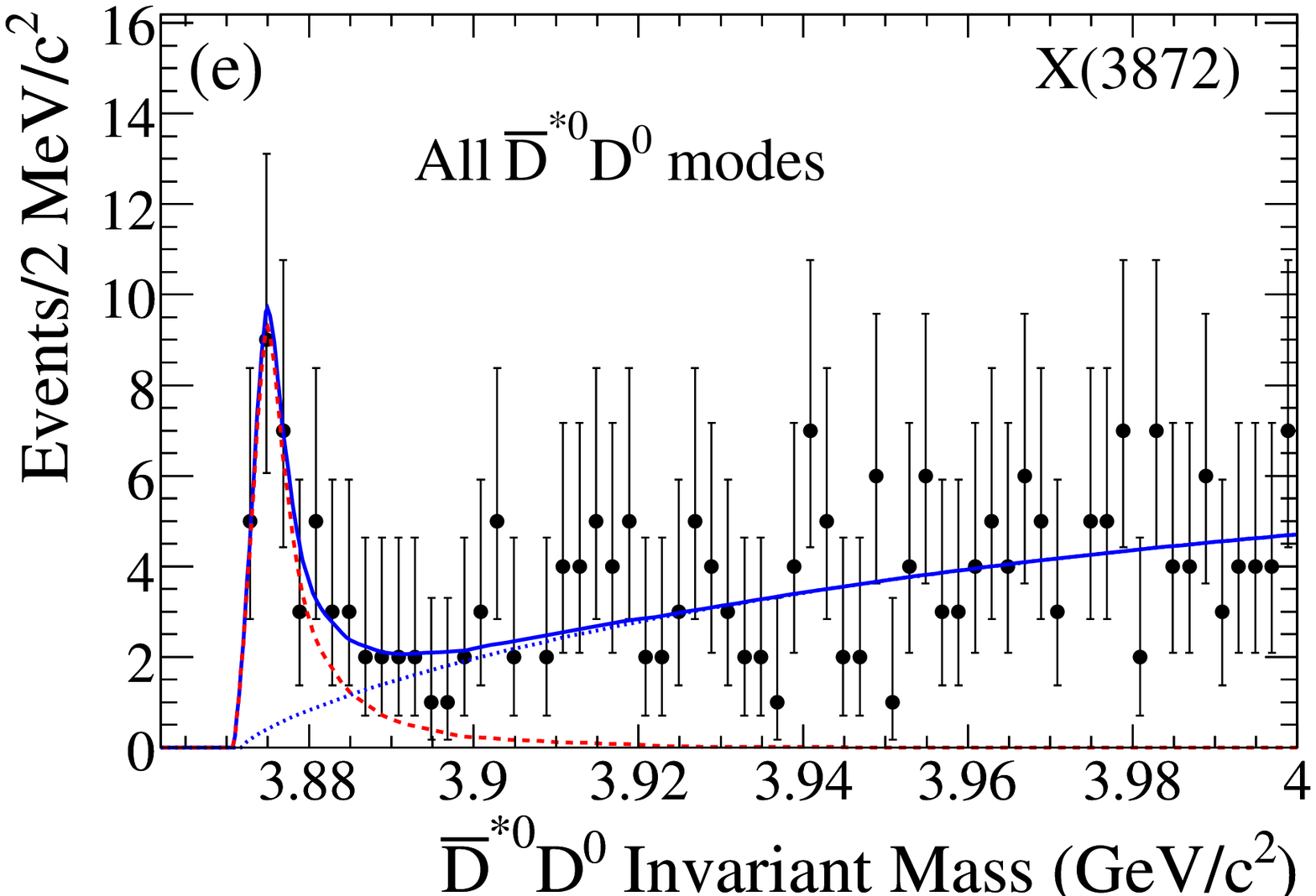}}
    \caption{The $D^0 \overline{D}^{(*)0} (\pi^0)$ invariant mass distributions in the $X(3872)$ region for selected $B$ decay events: a) Belle: $B^+ (B^0) \rightarrow D^0 \overline{D}^0 \pi^0 K^+ (K^0_\mathrm{S})$ ; b)\babar:$B^+ (B^0) \rightarrow D^0 \overline{D}^{*0} K^+ (K^0_\mathrm{S})$.}
    \label{fig:X3872-DDpi}
\end{figure*}

\begin{figure}[!h]
  \centering
    {\includegraphics[width=80mm]{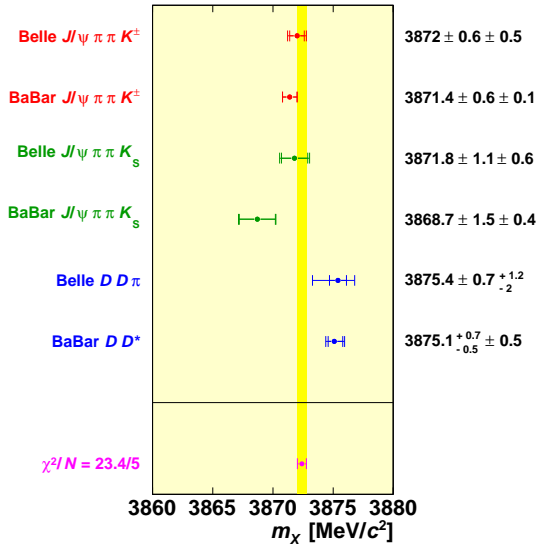}}
  \caption{Comparison of mass measurements for the $X(3872)$ state at $B$-factories.}
    \label{fig:X3872_massComp}
\end{figure}

Belle's observation of the $X(3872) \rightarrow D^0 \overline{D}^0 \pi^0$ decay mode has been more recently corroborated by a similar study by \babar~\cite{ref:BaBar_X3872_DDstar}. In a sample corresponding to 347 \invfb, $B^+$ ($B^0$) candidates have been reconstructed in the $B^+ (B^0) \rightarrow D^0 \overline{D}^{*0} K^+ (K^0_\mathrm{S})$ decay mode, where the $D^0$ are reconstructed in the $D^0 \rightarrow K^- \pi^+, K^- \pi^+ \pi^0, K^- \pi^+ \pi^- \pi^+$ channels, and combined with $\pi^0$ and $\gamma$ to give $D^{*0}$ candidates. The distribution of the $D^0 \overline{D}^{*0}$ invariant mass is shown in Fig.~\ref{fig:X3872-DDpi}, along with the $D^0 \overline{D}^0 \pi^0$ invariant mass distribution from Belle's analysis.\\
The measured value for the mass of the resonance is consistent with Belle's measurement; however, it is significantly higher than the average of the measurements from $X(3872) \rightarrow J/\psi~ \pi^+ \pi^-$ decays (Fig.~\ref{fig:X3872_massComp}). This might indicate the presence of two different states, as foreseen in the tetraquark model~\cite{ref:X3872_tetraquark}, decaying predominantly to different final states. A simple alternative explanation has been proposed in \cite{ref:X3872_massShift}, exploiting the fact that, because of the proximity of the $D^0 D^{*0}$ threshold, the peak position in the $m(D^0 \overline{D}^{*0})$ spectrum is sensitive to the orbital angular momentum of the final state.

\subsection{New states at 3940 \MeVcc}

\begin{figure}[h]
  \centering
    {\includegraphics[width=80mm]{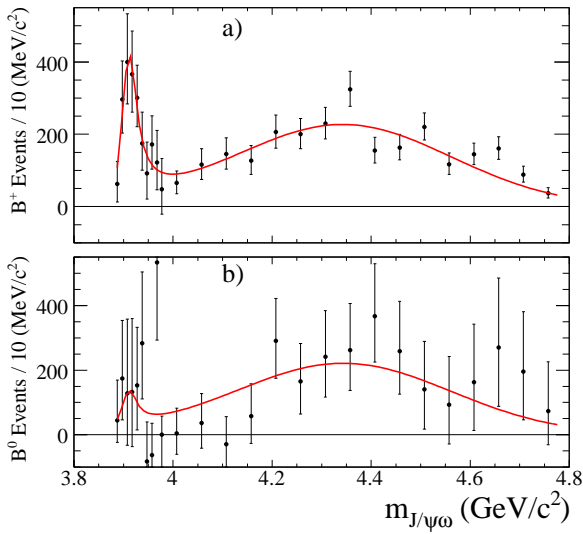}}
  \caption{\babar: the $J/\psi~ \omega$ invariant mass spectrum for selected $B^+ \rightarrow J/\psi~ \omega K^+$ (above) and $B^0 \rightarrow J/\psi~ \omega K^0_\mathrm{S}$ (below) events. Data in the neutral channel have been normalized to take into account the missing $K_\mathrm{L}$ and $K_\mathrm{S} \rightarrow \pi^0 \pi^0$ contributions. The superimposed curves show the result of the fit.}
    \label{fig:X3940_BaBar}
\end{figure}

In the mass region around 3940 \MeVcc, Belle has reported the observation of three resonant structures, provisionally named $X(3940)$, $Y(3940)$ and $Z(3930)$. The $X(3940)$ has been seen as a peak in the recoiling mass spectrum against a reconstructed $J/\psi$ produced in ``continuum'' \epem interactions; its decay to $D \overline{D}^*$ has been subsequently identified~\cite{ref:Belle_X3940}. The $Y(3940)$ appears as a threshold enhancement in the $J/\psi~ \omega$ invariant mass in reconstructed $B \rightarrow J/\psi~ \omega K$ decays~\cite{ref:Belle_Y3940}. Finally, the $Z(3930)$ has been seen in the $D \overline{D}$ invariant mass spectrum in selected $\gamma \gamma$ decays~\cite{ref:Belle_Z3930}.\\
Despite the proximity of the measured masses (see Table~\ref{tab:summary}), the structures are usually associated to three different states, mainly because of the different production and decay mechanism in which they have been observed. The mass and width values, the observed decay channel and a dedicated angular analysis make the $Z(3930)$ a good candidate for the $2^3P_2$ state of charmonium, the $\chi_{c2}(2P)$. For the other two states, proposed charmonium assignments include the $\eta_c(3S)$ and $\chi_{c1}(2P)$ for the $X(3940)$, and the $\chi_{c1}(2P)$ and $\chi_{c0}(2P)$ for the $Y(3940)$~\cite{ref:Eichten_review08}. However, these assignments lying on much weaker bases, different, unconventional interpretations (such as tetraquarks) have been proposed in this case too.\\
\babar~ has recently confirmed the observation of the $Y(3940)$ in a similar study of the $B^+ \rightarrow J/\psi~ \omega K^+$ and $B^0 \rightarrow J/\psi~ \omega K^0_\mathrm{S}$ decay modes~\cite{ref:BaBar_Y3940}, conducted on a sample corresponding to 348~\invfb. Combinations of $\pi^+ \pi^- \pi^0$ are selected if their invariant mass lies in the $\omega$ mass region, and coupled to reconstructed $J/\psi$ and $K^+$ ($K^0_\mathrm{S}$) to form $B^+$ ($B^0$) candidates. The helicity angle between the $\pi^+$ and the $\pi^0$ directions in the $\pi^+ \pi^-$ rest frame is then used to apply a weighting procedure in order to separate the $\omega$ contribution from the $\pi^+ \pi^- \pi^0$ combinatorial component: this establishes that practically all selected $B$ mesons decay to a real $\omega$. It is thus possible to ignore the shape of the $\pi^+ \pi^- \pi^0$ invariant mass distribution (which, in the $m(J/\psi~ \pi^+ \pi^- \pi^0)$ region of interest, is distorted due to the proximity of the production threshold), and to extract the number of $B$ signal events in different bins of the $J/\psi~ \omega$ invariant mass (Fig.~\ref{fig:X3940_BaBar}). A fit to the resulting distribution with the sum of a resonant and a smooth background shape yields the values of mass and width reported in Table~\ref{tab:summary}, both lower and significantly more precise than those obtained by Belle. The measured ratio of the neutral to charged $B$ decay rate to $Y K$ is $R_Y = 0.27^{+0.28}_{-0.23}~^{+0.04}_{-0.01}$.

\begin{table*}[!t]\addtolength{\tabcolsep}{1mm}
  \begin{center}
    \caption{Summary of the properties of recently discovered states in the charmonium region.\\}    \begin{tabular}{c | r @{ $\rightarrow$ } l l l | c @{ $\pm$ } c @{ $\pm$ } c c @{ $\pm$ } c @{ $\pm$ } c c}
      \hline \hline
      State & \multicolumn{2}{c}{Production mode} & Decay mode & Experiments & \multicolumn{3}{c}{$M (\MeVcc)$} & \multicolumn{3}{c}{$\Gamma (\MeVcc)$} & $J^{PC}$ \\
      \hline
      \multirow{2}{*}{$X(3872)$} & $B$ & $X K$ & \multirow{2}{*}{$J/\psi \pi^+ \pi^-, J/\psi \gamma$} & Belle, \babar, & \multicolumn{3}{l}{\multirow{2}{*}{$3871.2 \pm 0.5$}} & \multicolumn{3}{c}{\multirow{2}{*}{$< 2.3$}} & \multirow{2}{*}{$1^{++},$} \\
      & $p \overline{p}$ & $X \mathrm{[anything]}$ & & CDF, D0 & \multicolumn{3}{c}{} & \multicolumn{3}{c}{} & $2^{-+}$ \\
      $X(3875)$ & $B$ & $X K$ & $D^0 \overline{D}^0 \pi^0, D^0 \overline{D}^{*0}$ & Belle, \babar & 3875.2 & \multicolumn{2}{l}{0.7} & \multicolumn{2}{l @{ $\pm$ }}{$3.0~^ {+1.9}_{-1.4}$} & 0.9 \\
      \hline
      $X(3940)$ & $\epem$ & $J/\psi X$ & $D \overline{D}^*$ & Belle & 3943 & 6 & 6 & \multicolumn{3}{c}{$< 52$} & $J^{P+}$ \\
      $Y(3940)$ & $B$ & $Y K$ & $J/\psi \omega$ & Belle, \babar & \multicolumn{2}{l @{ $\pm$ }}{$3914.6~^{+3.8}_{-3.4}$} & 1.9 & \multicolumn{2}{l @{ $\pm$ }}{$34~^{+12}_{-8}$} & 5 & $J^{P+}$ \\
      $Z(3930)$ & $\gamma \gamma$ & $Z$ & $D \overline{D}$ & Belle & 3929 & 5 & 2 & 29 & 10 & 2 & $2^{++}$ \\
      \hline
      $Y(4008)$ & \epem & $Y$ & $J/\psi \pi^+ \pi^-, J/\psi \pi^0 \pi^0$ & Belle & 4008 & \multicolumn{2}{l}{$40~^{+114}_{-28}$} & 226 & 44 & 87 & $1^{--}$ \\
      $Y(4260)$ & \epem & $Y$ & $J/\psi \pi^+ \pi^-$ & \babar, Cleo, Belle & 4247 & \multicolumn{2}{l}{$12~^{+17}_{-32}$} & 108 & 19 & 10 & $1^{--}$ \\
      $Y(4350)$ & \epem & $Y$ & $\psi(2S) \pi^+ \pi^-,$ & \babar, Belle & 4361 & 9 & 9 & 74 & 15 & 10 & $1^{--}$ \\
      $Y(4660)$ & \epem & $Y$ & $\psi(2S) \pi^+ \pi^-$ & Belle & 4664 & 11 & 5 & 48 & 15 & 3 & $1^{--}$ \\
      \hline
      $Y(4160)$ & \epem & $J/\psi Y$ & $D^* \overline{D}^*$ & Belle & 4433 & 4 & 2 & \multicolumn{3}{l}{$44~^{+18}_{-13}~^{+30}_{-13}$} & $J^{P+}$ \\
      \hline
    \end{tabular}
  \label{tab:summary}
  \end{center}
\end{table*}

\subsection{The $Y$ vector states}

\begin{figure*}[!t]
  \centering
    \subfigure[]
      {\includegraphics[width=75mm]{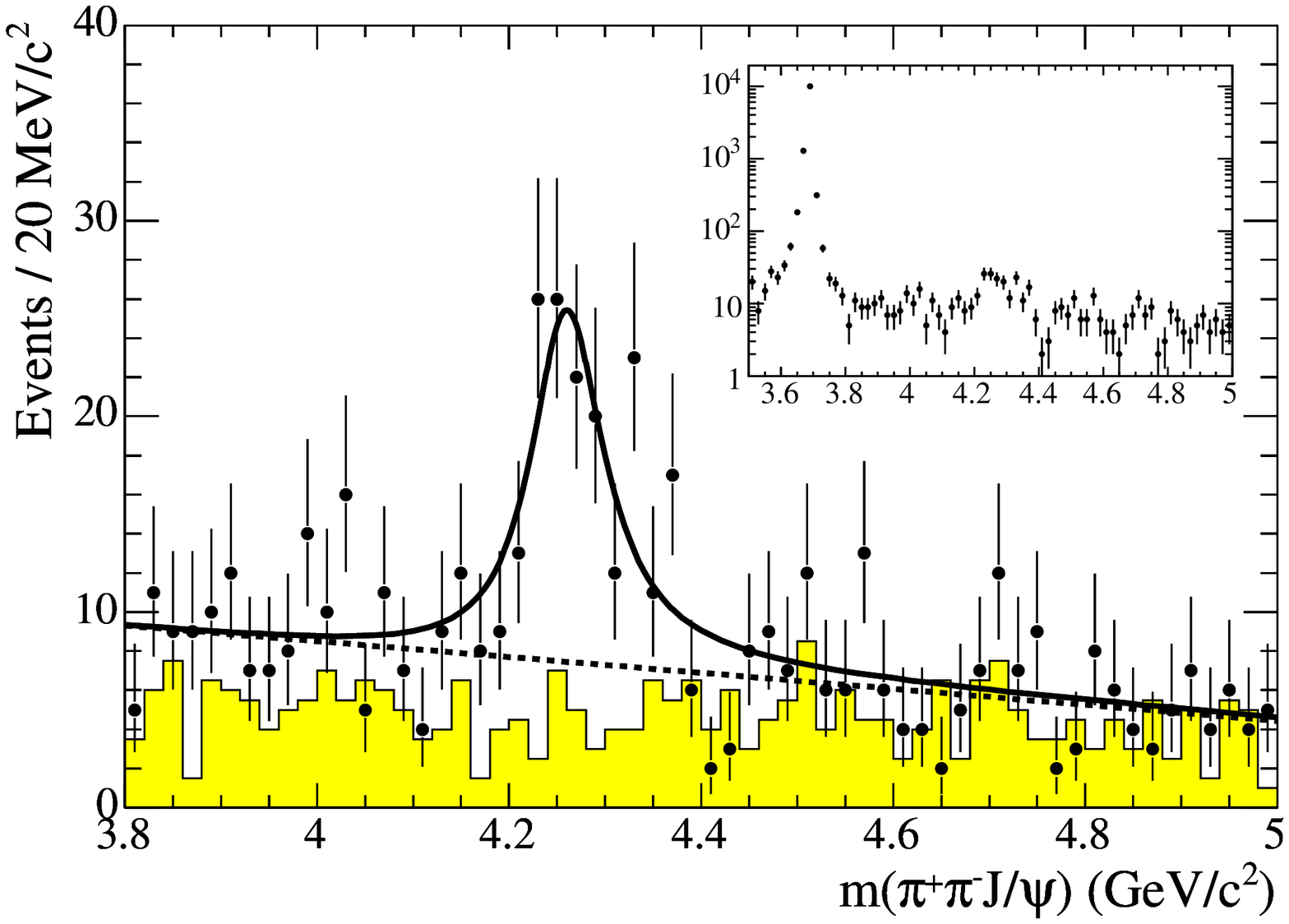}}
    \subfigure[]
      {\includegraphics[width=85mm]{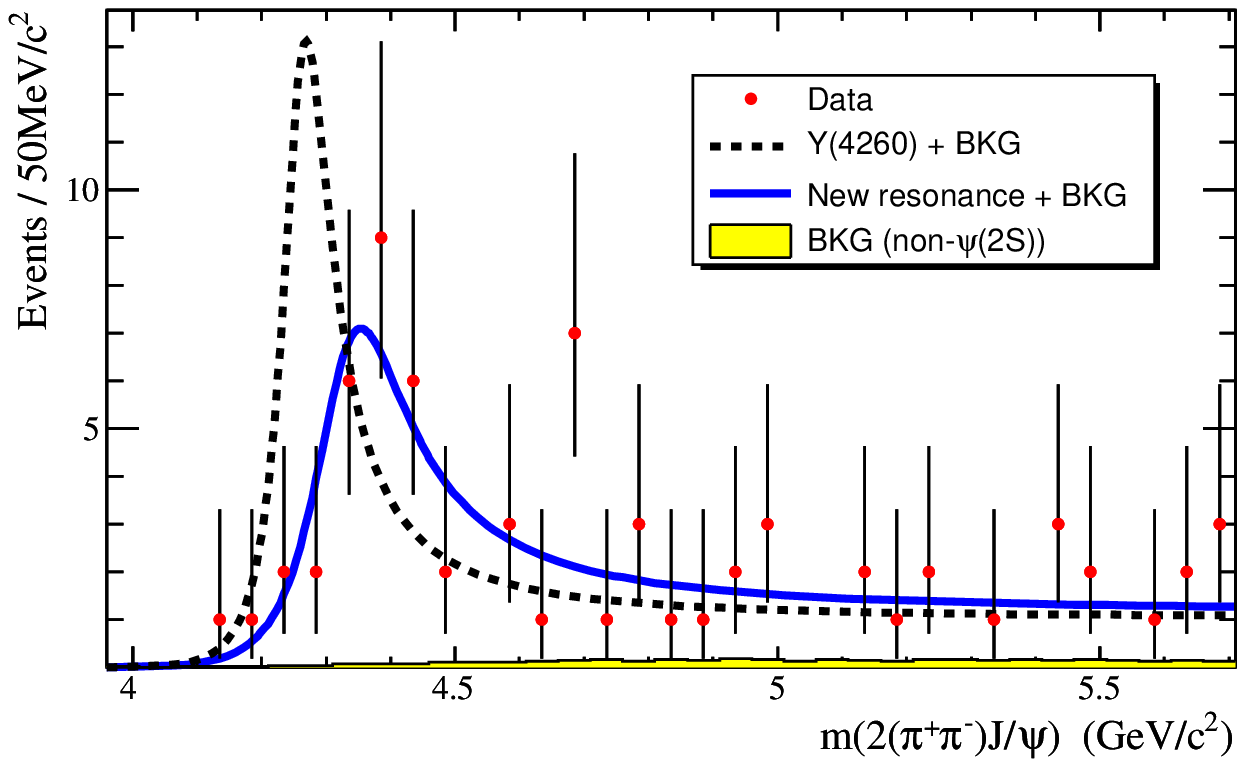}}
    \subfigure[]
      {\includegraphics[width=78mm]{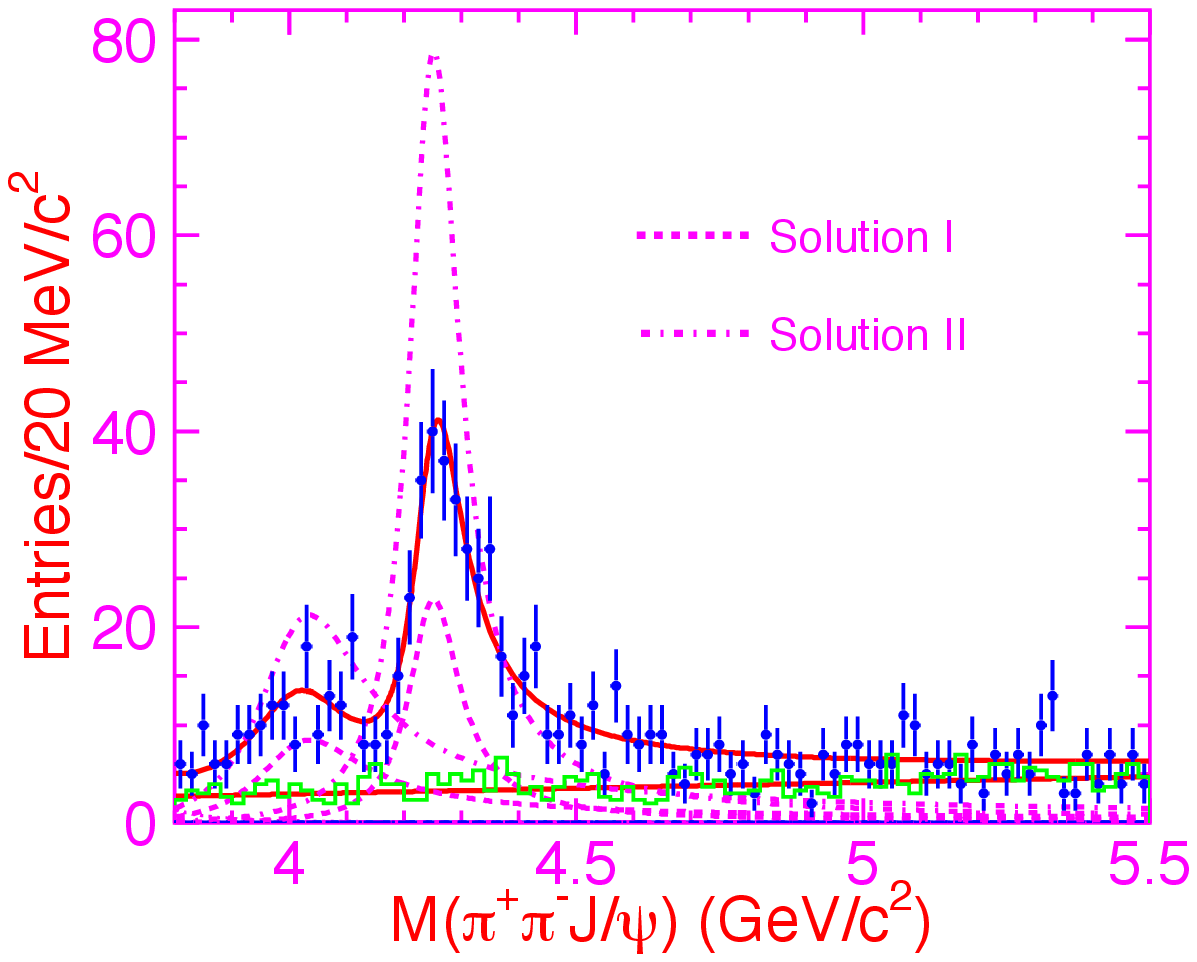}}
      \hspace{4mm}
    \raisebox{4mm}{\subfigure[]
      {\includegraphics[width=78mm]{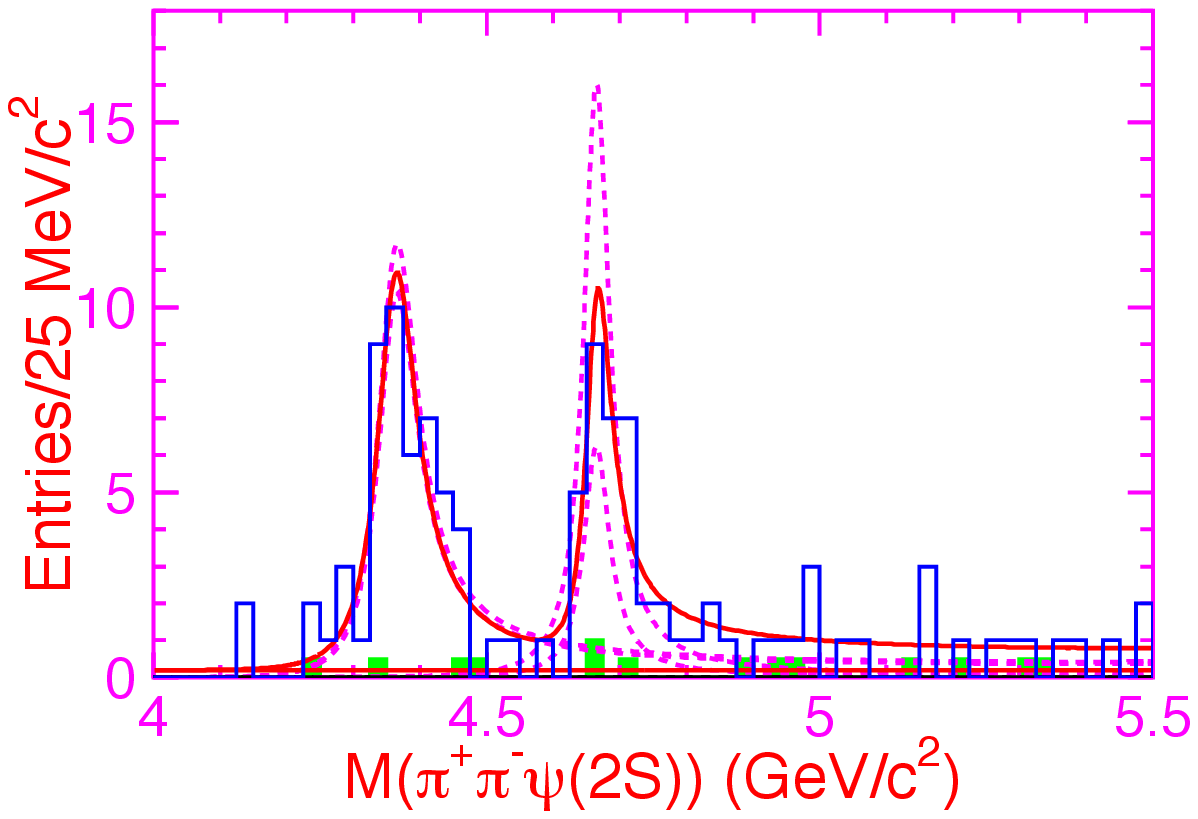}}}
      \hspace{4mm}
    \caption{Evidence for the vector `$Y$' states in ISR events at $B$-factories. Spectra for final state invariant masses are shown: a) \babar, $m(J/\psi~ \pi^+ \pi^-)$; b) \babar, $m(\psi(2S)~ \pi^+ \pi^-)$; c) Belle, $m(J/\psi~ \pi^+ \pi^-)$; d) Belle, $m(\psi(2S)~ \pi^+ \pi^-)$.}
    \label{fig:Ystates_BaBar-Belle}
\end{figure*}

Another set of unexpected structures in the mass region above the open charm production threshold has been revealed by the study of ISR events where a $J/\psi$ or a $\psi(2S)$ is produced in the final state.\\
The first of these states has been observed by \babar~ \cite{ref:BaBar_Y4260} as a broad enhancement around $4.26~\GeVcc$ in the spectrum of the $J/\psi~\pi^+ \pi^-$ invariant mass (Fig.~\ref{fig:Ystates_BaBar-Belle}a) in ISR events $\epem \rightarrow J/\psi~\pi^+ \pi^- \gamma_\mathrm{ISR}$. The selection required no other charged particle detected except those reconstructing the final state; the detection of the ISR photon, preferentially emitted along the beam direction, was not required. Moreover, the total transverse momentum and the inferred recoiling mass were required to be consistent with the kinematics of the ISR process. The provisional name given to the resonance is $Y(4260)$.\\
In a similar analysis performed on a sample of $13.3~\invfb$ of data collected at energies between the \YOneS~ and \YFourS~ peaks, CLEO has been able to confirm the observation of the $Y(4260)$~\cite{ref:CLEO_Y4260}. An independent confirmation has come from an energy scan performed by CLEO-c in the region $3.77~\mathrm{GeV} < \sqrt{s} < 4.26~\mathrm{GeV}$, where the exclusive reaction $\epem \rightarrow J/\psi~ \pi^+ \pi^-$ has been reconstructed. In the same analysis, a significant enhancement at $\sqrt{s} = 4.26~\mathrm{GeV}$ has also been observed in the $J/\psi~ \pi^0 \pi^0$ final state, thus providing evidence for the $Y(4260) \rightarrow J/\psi~ \pi^0 \pi^0$ decay mode \cite{ref:CLEO-c_Y4260}.\\
By repeating on the $\psi(2S)~ \pi^+ \pi^-$ final state the same kind of analysis leading to the observation of the $Y(4260)$, \babar~ has identified another broad structure just above the production threshold, at a mass around $4.32~\GeVcc$ (Fig.~\ref{fig:Ystates_BaBar-Belle}b)~\cite{ref:BaBar_Y4350}.\\
More recently, Belle has repeated the same studies with much higher statistics. On a data sample corresponding to $548~\invfb$, ISR events with $J/\psi~\pi^+ \pi^-$ final state have been reconstructed and the corresponding invariant mass spectrum analysed~\cite{ref:Belle_Y4008}. Beside the clear peak of the $Y(4260)$ (Fig.~\ref{fig:Ystates_BaBar-Belle}c), another broad enhancement is visible, which Belle fits with a resonant shape and names $Y(4008)$. On a larger sample ($673~\invfb$), the study has been repeated for the  $\psi(2S)~\pi^+ \pi^-$ final state~\cite{ref:Belle_Y4660}: the analysis of the resulting spectrum confirms with high significance the presence of the $Y(4350)$ peak (Fig.~\ref{fig:Ystates_BaBar-Belle}d), and reveals yet another broad structure, at a mass around $4.66~\GeVcc$. The mass and width of all `$Y$' states, as obtained by Belle from fits to Breit-Wigner resonant shapes, are reported in Table~\ref{tab:summary}.\\
All these states share several properties: they are vector states with the photon quantum numbers, $J^{PC} = 1^{--}$; they have large total widths; they have only been observed in charmonium decay modes. Not only they do not correspond to any known charmonium state, but were they all real and distinct resonances, they would clearly overpopulate that mass region. The most favourable charmonium assignment for the $Y(4260)$ would be the $\psi(4S)$~\cite{ref:Y4260_psi4S}, but the $\psi(4415)$ has already been proposed as a candidate; moreover, a preliminary upper limit of 7.6 found for the ratio of the $D \overline{D}$ to $J/\psi~\pi^+ \pi^-$ decay rate appears far too low for this assignment. Alternative explanations include tetraquark objects~\cite{ref:Y4260_tetraquark} and $c \overline{c} g$ hybrids~\cite{ref:Y4260_hybrid}.

\subsection{Overview}

Table~\ref{tab:summary} summarizes the main properties of the states recently discovered at $B$-factories in the charmonium region (it also includes the $Y(4160)$, seen by Belle in double-charmonium production and not discussed in this work). Except for the $Z(3930)$, which is likely to be the $\chi_{c2}(2P)$ state of charmonium, it is quite hard to fit them into the spectrum of conventional $c \overline{c}$ mesons. Some of them might in fact represent the first clear indication of a new non-conventional hadron spectroscopy.

\section{The bottomonium region}

\subsection{Bottomonium spectroscopy at $B$-factories}

The $B$-factories are not usually considered as ideal facilities for the study of the bottomonium spectrum, their energy being tuned to the peak of the \YFourS~resonance, which decays in almost 100\% of cases to a $B\overline{B}$ pair.\\
However, the huge amount of data accumulated by \babar~ and Belle in several years of running allows for the search for rare, non-$B\overline{B}$ decay modes. Moreover,  the ISR mechanism provides a way to access directly the states below the open-$B$ threshold, making for some of the largest data sets ever recorded in this region.\\
Another important option is given by the possibility of running the \epem~ colliders at energies different from the \YFourS~mass. In fact, both \babar~ and Belle have recorded, in short periods of time, data samples at various energies in the bottomonium region which outrun all pre-existing sample in that range. They are summarized in Table~\ref{tab:nonY4S}.\footnote{A few weeks after this Conference, from an analysis of the data sample recorded at the \YThreeS~peak, \babar~has reported the first observation of the ground state of bottomonium, the $\eta_b(1S)$.}

\begin{table}[!h]\addtolength{\tabcolsep}{2mm}
  \begin{center}
    \caption{\babar~ and Belle data samples away from the \YFourS~ peak.\\}
    \begin{tabular}{l | c c}
      \hline \hline
      & Energy & $\int{{\cal L} \mathrm{d}t}~ (\invfb)$ \\
      \hline
      \multirow{3}{*}{\babar} & \YTwoS~ peak & 15 \\
      & \YThreeS~ peak & 30 \\
      & $\YFourS - 11.3~\GeVcc$ & 11.3 \\
      \hline
      \multirow{3}{*}{Belle} & \YThreeS~ peak & 3 \\
      & \YFiveS~ scan & 2 \\
      & \YFiveS~peak & 21.7 \\
      \hline
    \end{tabular}
  \label{tab:nonY4S}
  \end{center}
\end{table}

\subsection{Hadronic decays of the \YFourS}

\begin{figure*}[bh!]
  \centering
    \subfigure[]
      {\includegraphics[width=80mm]{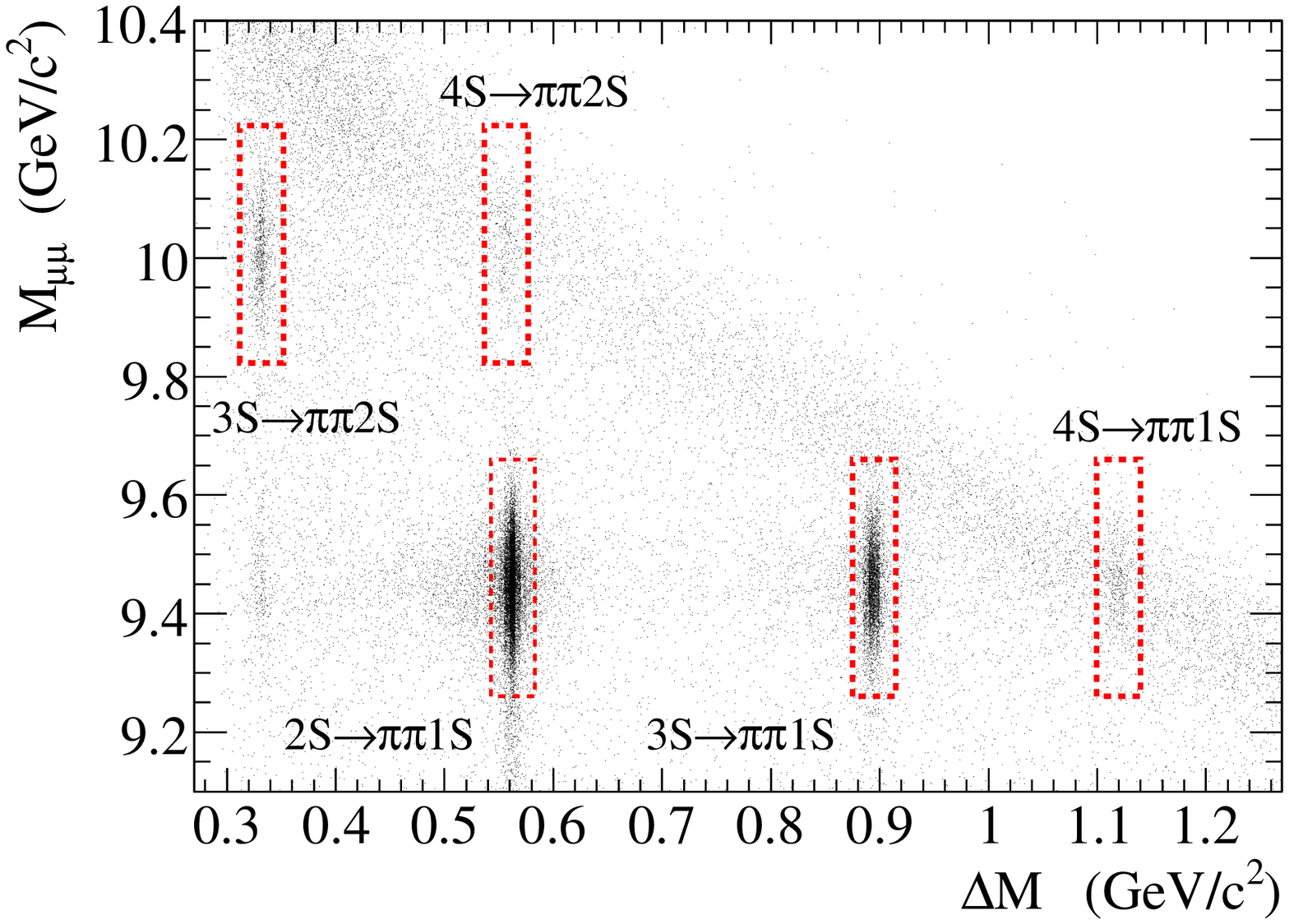}}
    \raisebox{8mm}{\subfigure[]
      {\includegraphics[width=44mm]{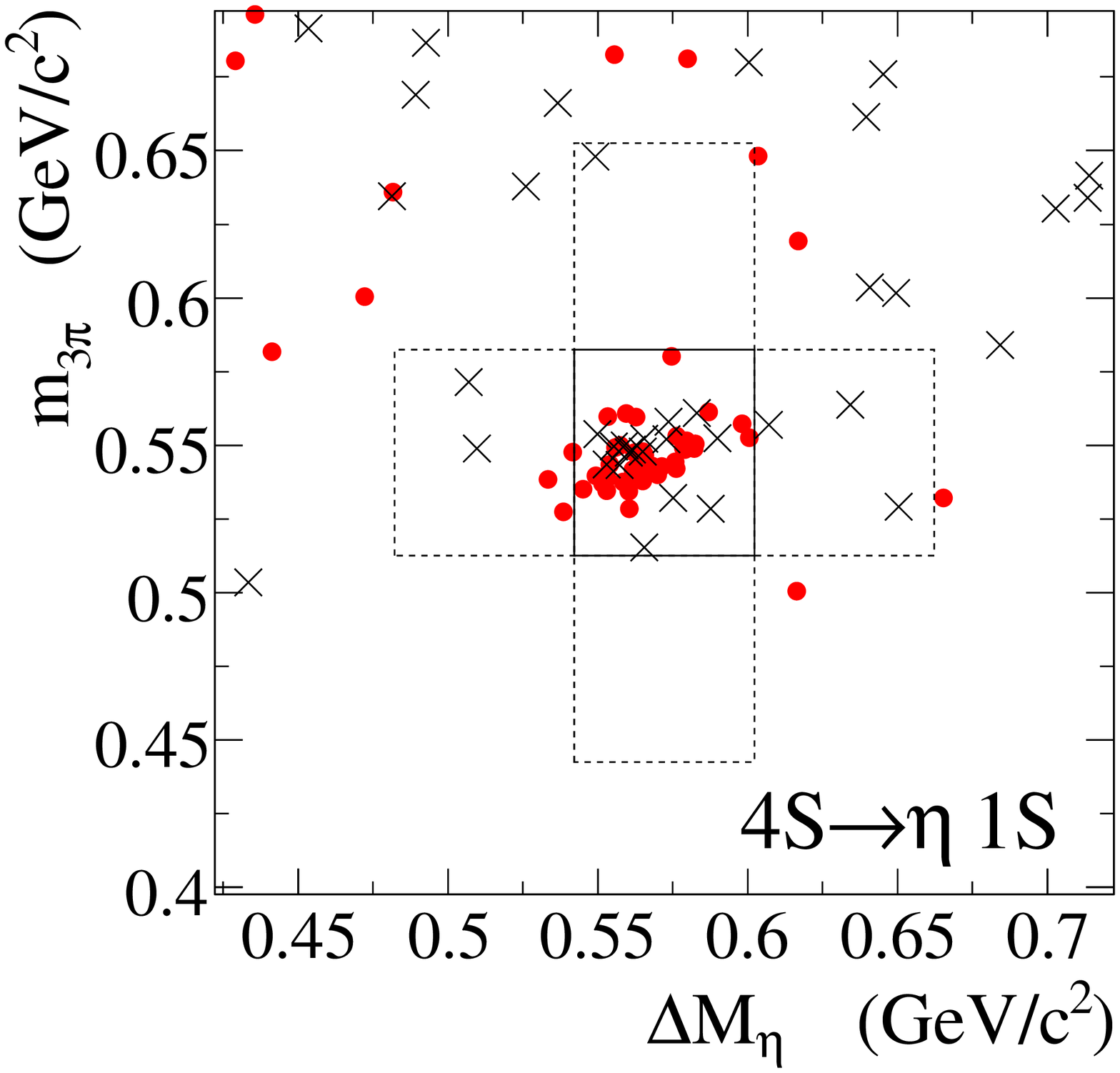}}}
    \raisebox{8mm}{\subfigure[]
      {\includegraphics[width=44mm]{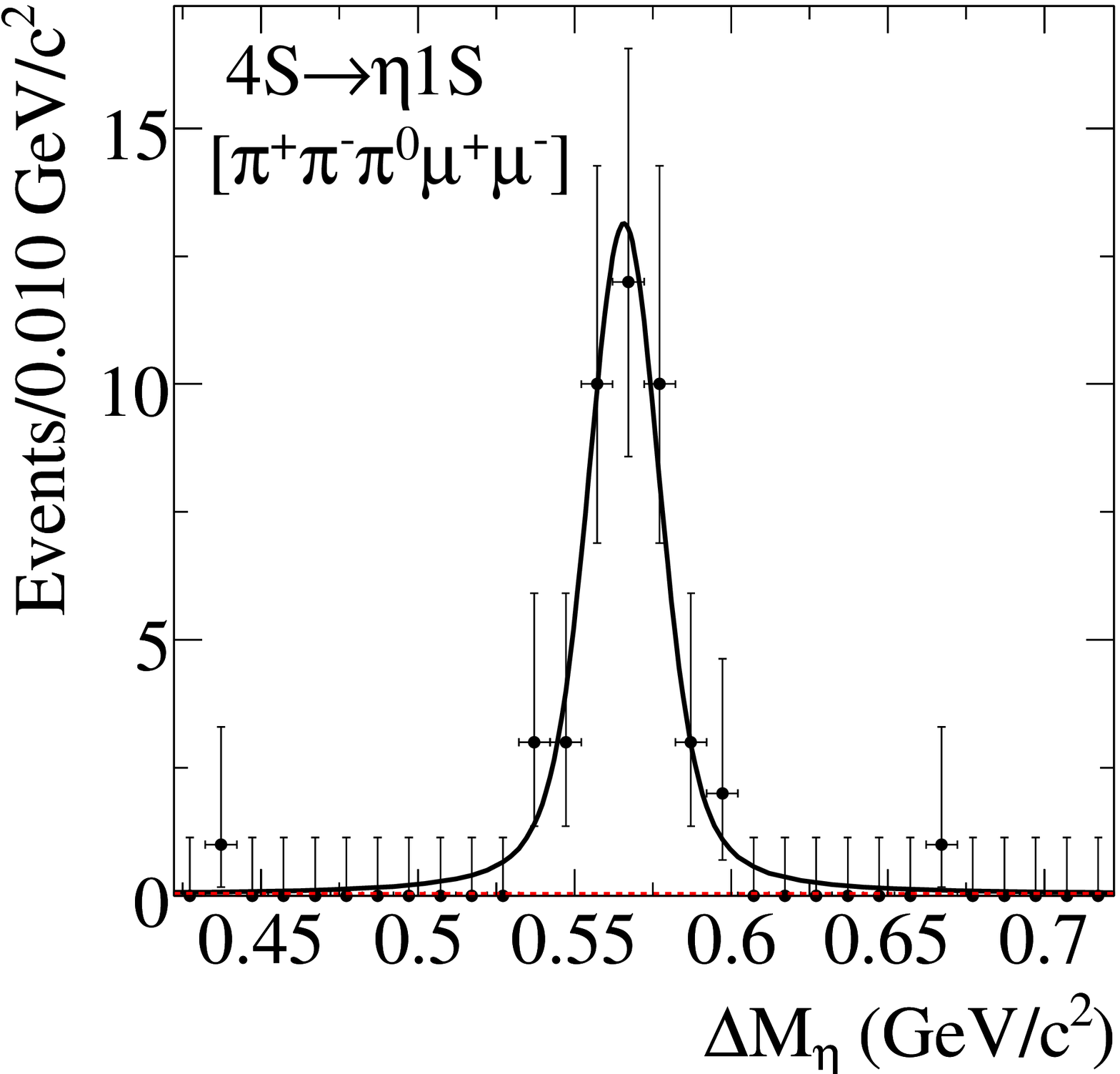}}}
    \caption{Study of $\Upsilon(nS)$ hadronic decay modes at \babar: a)scatter plot of $m(\mu^+ \mu^-)$ vs. $\Delta M_{\pi\pi}$ for selected $\mu^+ \mu^- \pi^+ \pi^-$ events: the boxes indicate the different signal regions; b) distribution of $m(\pi^+ \pi^- \pi^0)$ vs. $\Delta M_\eta$ for selected $\mu^+ \mu^- \pi^+ \pi^- \pi^0$ (red circles) and $\epem \pi^+ \pi^- \pi^0$ (black crosses) events: the boxes indicate the signal regions in the two variables; c) projection of the $\Delta M_\eta$ distribution for $\mu^+ \mu^-$ events in the $m(\pi^+ \pi^- \pi^0)$ signal region in (b).}
    \label{fig:Y4S-etaY1S}
\end{figure*}

Study of the hadronic transitions between bottomonium states provides a way to test QCD multipole expansion (QCDME) models, which are usually employed to describe this kind of process.\\
Observation of non-$B \overline{B}$ hadronic decays of the \YFourS~ has been first published by \babar~\cite{ref:BaBar_Y4S-pipiYnS}, following evidence for the $\Upsilon(4S) \rightarrow \Upsilon(1S)~ \pi^+ \pi^-$ decay mode previously reported by Belle; all results have been confirmed by a more recent update from Belle~\cite{ref:Belle_Y4S-pipiYnS}. In these analyses, events with a muon pair $\mu^+ \mu^-$ and a charged pion pair $\pi^+\pi^-$ were selected. Then the distribution of $m(\mu^+ \mu^-)$ vs. $\Delta M_{\pi\pi} \equiv m(\mu^+ \mu^- \pi^+ \pi^-) - m(\mu^+ \mu^-)$ was studied in order to identify decay chains of the type $\Upsilon(mS) \rightarrow \Upsilon(nS)~ \pi^+ \pi^-, \Upsilon(nS) \rightarrow \mu^+ \mu^-$. In addition to $\YThreeS \rightarrow \Upsilon(1S, 2S)~ \pi^+ \pi^-$ and $\YTwoS \rightarrow \YOneS~ \pi^+ \pi^-$ events from ISR production, used as control samples, significant signals for $\YFourS~ \rightarrow \YOneS~ \pi^+ \pi^-$ and $\YFourS \rightarrow \YTwoS~ \pi^+ \pi^-$ were found, with branching fractions of the order of $10^{-3}$.\\
In a recent update~\cite{ref:BaBar_Y4S-etaY1S}, \babar~ has used a data sample corresponding to 348~\invfb and added the study of the $\ell^+ \ell^- \eta$ final state, where the $\eta$ is reconstructed in the $\eta \rightarrow \pi^+ \pi^- \pi^0$ channel. Due to the increased statistics, the \epem~ sample now provides significant signals in many of the modes studied.\\
Figure~\ref{fig:Y4S-etaY1S}a shows the distribution of $m(\mu^+ \mu^-)$ vs. $\Delta M_{\pi\pi}$ for selected $\mu^+ \mu^- \pi^+ \pi^-$ events. After a further selection, the signal yield for each of the decay modes is obtained from fits to the $\Delta M_{\pi\pi}$ distributions for events within appropriate windows in $m(\ell^+ \ell^-)$. Extraction of the branching fractions provides the most precise determination to date of $\BR(\YFourS~ \rightarrow \Upsilon(1S, 2S)~ \pi^+ \pi^-)$ and $\BR(\YThreeS~ \rightarrow \YTwoS~ \pi^+ \pi^-)$ and measurements of $\BR(\Upsilon(3S, 2S) \rightarrow \YOneS~ \pi^+ \pi^-)$ comparable to current world averages.\\
Figure~\ref{fig:Y4S-etaY1S}b shows the distribution of $m(\pi^+ \pi^- \pi^0)$ vs. $\Delta M_\eta \equiv m(\ell^+ \ell^- \pi^+ \pi^- \pi^0) - m(\ell^+ \ell^-) - m(\pi^+ \pi^- \pi^0)$ for selected $\ell^+ \ell^- \pi^+ \pi^- \pi^0$ events. In this case signal yields are obtained from fits to the $\Delta M_\eta$ distributions for events within an appropriate window in $m(\pi^+ \pi^- \pi^0)$ (Fig.~\ref{fig:Y4S-etaY1S}c). Significant yields ($11 \sigma$ and $6.2 \sigma$, respectively) are obtained in both the $\mu^+ \mu^-$ and \epem~ samples, thus providing the first observation of the $\YFourS \rightarrow \YOneS~ \eta$ decay mode. The resulting branching fraction, $\BR(\YFourS \rightarrow \YOneS~ \eta) = 1.96 \pm 0.06 \pm 0.09$, is about 2.4 times larger than the corresponding dipion rate.

\subsection{Anomalous \YFiveS~decay rates} 

\begin{figure}[!h]
  \centering
    \subfigure[]
      {\includegraphics[width=80mm]{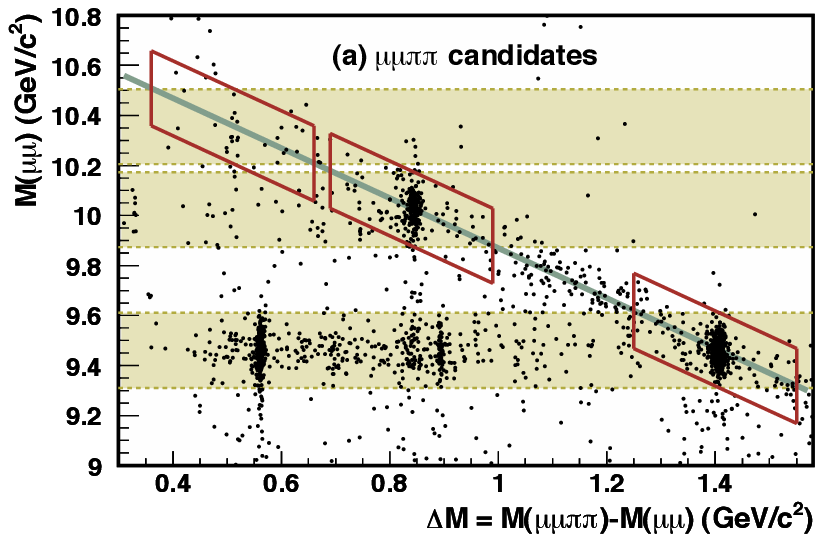}}
    \\
    \hspace{1mm}
    \subfigure[]
      {\includegraphics[width=78mm]{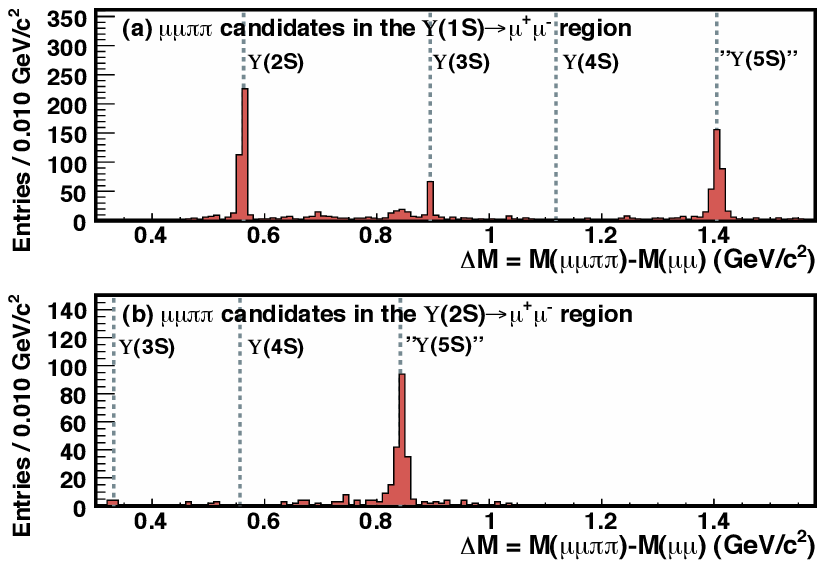}}
    \caption{Study of $\Upsilon(nS)~ \pi^+ \pi^-$ final states in Belle's $\Upsilon(5S)$ sample: a) scatter plot of $m(\mu^+ \mu^-)$ vs. $\Delta M_{\pi\pi}$ for selected $\mu^+ \mu^- \pi^+ \pi^-$ events: horizontal bands indicate the \YOneS, \YTwoS~ and \YThreeS~ signal regions; b) projected distributions in $\Delta M_{\pi\pi}$ for events in the \YOneS~ (above) and \YTwoS~ (below) bands in (a).}
    \label{fig:anomYnSpipi}
\end{figure}

(Note: this topic is discussed in much more detail in Ref.~\cite{ref:Wicht_HQL08} in these Proceedings.)\\
Another study of hadronic $\Upsilon(mS) \rightarrow \Upsilon(nS)~ \pi^+ \pi^-$ transitions has been performed by Belle, using the data sample recorded at $\sqrt{s} \sim 10.87~\mathrm{GeV}$, near the peak of the \YFiveS. Fig.~\ref{fig:anomYnSpipi}a shows the distribution of $m(\mu^+ \mu^-)$ vs. $\Delta M_{\pi\pi}$ for selected $\mu^+ \mu^- \pi^+ \pi^-$ events; the horizontal bands indicate the \YOneS, \YTwoS~ and \YThreeS~ signal regions (bottom to top) for $m(\mu^+ \mu^-)$.\\
Fig.~\ref{fig:anomYnSpipi}b shows the $\Delta M_{\pi\pi}$ projections for events in the \YOneS~ and \YTwoS~ band, where clear peaks corresponding to the $\YThreeS \rightarrow \YOneS~ \pi^+ \pi^-$, $\YTwoS \rightarrow \YOneS~ \pi^+ \pi^-$ and $\YFiveS \rightarrow \YOneS~ \pi^+ \pi^-$, $\YFiveS \rightarrow \YTwoS~ \pi^+ \pi^-$ transitions can be clearly seen. Similar peaks are found, with a somewhat smaller significance, in the $\YFiveS \rightarrow \YThreeS~ \pi^+ \pi^-$ and $\YFiveS \rightarrow \YOneS~ K^+ K^-$ modes.\\
However, when partial widths are extracted from the data, they result in values more than two orders of magnitude larger than the corresponding ones for \YTwoS, \YThreeS~ and \YFourS~ decays. This observation prompts the question whether the observed enhancement at $m(\epem) \sim 10.87~\GeVcc$ really corresponds to the bottomonium state \YFiveS, or is rather suggestive of a ``hidden beauty counterpart'' of the $Y(4260)$ state observed in the charmonium region, which shows a surprinsingly large decay rate to $J/\psi~ \pi^+ \pi^-$.

\section{Summary}

The $B$-factories have made a significant contribution to our knowledge of the charmonium states (observation of the $\eta_c$ and possibly of the $\chi_{2c}(2P)$, new measurements of resonance parameters); but most of all, in recent years, with the discovery of many unexpected and often puzzling states in the charmonium mass region, may have unveiled a whole new type of spectroscopy.\\
The quest for new discoveries is now being extended to the bottomonium sector, where thanks to the large amount of data accumulated and to dedicated run settings, the possibility of finding new exciting results appears concrete and is in fact already manifesting itself.

\bigskip 

\end{document}